\documentclass[journal]{IEEEtran} 

\usepackage{amssymb,amsthm,amsmath}
\usepackage{graphicx}                       
\usepackage{epstopdf}
\usepackage{siunitx}                        
\usepackage{subfigure}                      
\usepackage{fancyhdr}                       
\usepackage{balance}                        
\usepackage[hidelinks]{hyperref}            

\graphicspath{{../}{./figures/}}
\UseRawInputEncoding

\usepackage{booktabs}
\usepackage{threeparttable}
\usepackage{multirow}
\usepackage{url}
\usepackage{algorithm}
\usepackage[noend]{algpseudocode}
\usepackage{makecell}
\usepackage{multirow}

\usepackage{enumitem}
\usepackage{tabularray}
\usepackage{tabu}
\usepackage{booktabs}
\usepackage{stfloats}
\usepackage{cite}
\usepackage{cuted}

\usepackage{xcolor,soul,framed} 
\colorlet{shadecolor}{yellow}
\hypersetup{
colorlinks=true,
linkcolor=blue,
anchorcolor=red,
citecolor=blue
}

\begin{document}
\title{\LARGE{Integrated Power and Thermal Management for Enhancing Energy Efficiency and Battery Life in Connected and Automated Electric Vehicles}}
%
%
\author{Dongjun Li,
        Qiuhao Hu,
        Weiran Jiang,
	    Haoxuan Dong,
        and Ziyou Song 
\thanks{This work was supported in part by A-Star Young Individual Research Grants (YIRG), Singapore under Grant M22K3c0092. \textit{(Corresponding author: Ziyou Song.)}}
\thanks{D. Li, H. Dong, and Z. Song are with the Department of Mechanical Engineering, National University of Singapore, 117575 Singapore, Singapore (e-mail: dongjun.li@u.nus.edu; donghaox@foxmail.com; ziyou@nus.edu.sg).}
\thanks{Q. Hu and W. Jiang are with the Farasis Energy, Hayward, CA, 94545 USA (email: bhu@farasis.com; wjiang@farasis.com).}
}
%
%

\markboth{ }
{Shell \MakeLowercase{\textit{et al.}}: Bare Demo of IEEEtran.cls for IEEE Journals}



\maketitle 
\begin{abstract}
Effective power and thermal management are essential for ensuring battery efficiency, safety, and longevity in Connected and Automated Electric Vehicles (CAEVs). However, real-time implementation is challenging due to the multi-timescale dynamics and complex trade-offs between energy consumption, battery degradation, traffic efficiency, and thermal regulation. This paper proposes a novel integrated power and thermal management strategy based on the Multi-Horizon Model Predictive Control (MH-MPC) framework to enhance energy efficiency, optimize battery temperature, ensure traffic safety and efficiency, and reduce battery degradation for CAEVs. The proposed strategy is formulated with a focus on the aging term, allowing it to more effectively manage the trade-offs between energy consumption, battery degradation, and temperature regulation Moreover, the proposed strategy leverages multi-horizon optimization to achieve substantial improvements, reducing computation time by $7.18\%$, cooling energy by $14.22\%$, traction energy by $8.26\%$, battery degradation loss by over $22\%$, and battery degradation inconsistency by $36.57\%$ compared to the benchmark strategy. Furthermore, sensitivity analyses of key parameters, including weighting factors, sampling time, and prediction horizons, demonstrate the robustness of the strategy and underscore its potential for practical applications in extending battery lifespan while ensuring safety and efficiency.
\end{abstract}

%
\IEEEpeerreviewmaketitle

\section{Introduction}
%
%
%
%

The emergence of connected and automated vehicles (CAVs) presents a promising opportunity to improve energy efficiency and enhance traffic safety through advanced sensors and vehicle-to-everything (V2X) technology \cite{abboud2016interworking}. Current research on CAVs, such as eco-driving in mixed traffic flow, primarily focuses on reducing traction energy consumption by dynamically adjusting vehicle acceleration in response to real-time and predicted traffic information from surrounding vehicles \cite{li2024physics,li2024physicsinformed,lu2024safe}. However, few studies focus on simultaneously optimizing traction energy efficiency, battery thermal management, and battery degradation \cite{amini2020hierarchical,wu2024optimal,wu2024integrated}.

Battery thermal management systems (BTMS) are critical for ensuring the safety, efficiency, and longevity of battery packs, particularly in electric vehicles (EVs) \cite{kim2019review,li2021innovative,wu2019critical}. As the demand for EVs rises in response to the global shift towards sustainable transportation, optimizing BTMS performance becomes increasingly important. A well-designed BTMS helps maintain batteries within their optimal temperature range, enhancing safety and performance while prolonging battery life. 
Moreover, battery degradation is significantly influenced by the performance of BTMS and vehicle traction power, as factors like temperature and current interact to create complex feedback loops \cite{amini2020hierarchical,hajidavalloo2023nmpc}. Therefore, simultaneously optimizing power and thermal management is critical to ensuring the safety, traffic efficiency, energy efficiency, and lifespan of EVs.

There are several challenges in developing a real-time integrated power and thermal management (IPTM) strategy for connected and automated electric vehicles (CAEVs), particularly when considering battery degradation \cite{amini2020hierarchical,wu2024optimal}. These challenges stem from the following key factors:
\begin{enumerate}
    \item \textit{Multi-timescale dynamics:} The thermal dynamics of the battery pack change much more slowly than the vehicle's velocity and inter-vehicle spacing \cite{hu2023robust}. This delayed response can result in insufficient cooling during sudden high-power events, such as uphill driving or frequent acceleration and deceleration. Maintaining battery temperatures within the optimal range $(20^{o}\mathrm{C}$ to $30^{o}\mathrm{C})$ \cite{guo2024energy,kumar2024inverse} while minimizing energy consumption is difficult. The slow thermal response can cause overheating and fluctuations in cooling power, increasing energy consumption. Although extending prediction horizons and selecting appropriate sampling times can help balance multi-timescale dynamics, doing so significantly increases the computational footprint, which complicates real-time implementation \cite{park2020computationally}.
    \item \textit{Coupled Trade-offs}: Solving the optimization problem with multiple objectives involves navigating trade-offs between competing objectives, such as cooling energy efficiency versus battery degradation \cite{wu2024optimal}, traction energy efficiency versus traffic efficiency \cite{li2024physics}, and temperature regulation versus overall energy efficiency. For example, while lower battery temperatures reduce degradation, increased cooling energy elevates current, which in turn exacerbates degradation. Likewise, larger inter-vehicle spacing improves energy efficiency but reduces traffic throughput, whereas smaller spacing compromises energy efficiency and safety \cite{li2024physics}. 
\end{enumerate}

To address these challenges, several studies have proposed solutions. For example, Amini et al. \cite{amini2020hierarchical} introduced a hierarchical model predictive control (MPC) framework for robust eco-cooling in CAEVs, optimizing power and thermal management to improve energy efficiency. However, this study simplifies the thermal dynamics and does not account for battery degradation. Zhao et al. \cite{zhao2021two} developed a two-layer predictive control strategy that balances energy savings with an expanded control range for battery and cabin thermal management, achieving significant energy reductions. Similarly, Ma et al. \cite{ma2024two} further contributed by proposing a two-level optimization strategy for vehicle speed and battery thermal management, improving energy efficiency and battery safety. This study decouples vehicle speed and thermal control in the upper layer while coupling them in the lower layer, leading to substantial energy efficiency gains. While these BTMS-related studies successfully integrate power and thermal management to improve energy efficiency, they overlook the impact of battery degradation, which remains a critical area for further research.

In the context of battery degradation optimization, our previous work \cite{wu2024optimal} explored the trade-off between cooling energy efficiency and battery degradation using dynamic programming, demonstrating improvements in degradation loss by optimally timing the activation of the AC system. However, this study focuses exclusively on battery thermal management within a fixed-speed planning framework, limiting its scope to more comprehensive power and thermal management optimization. Additionally, the intricate dynamics in battery temperature and degradation have constrained the use of high-fidelity battery pack models, leading many studies to adopt simplified models that overlook inconsistencies within battery packs \cite{wei2024modeling}.

To address the aforementioned challenges, we propose a novel IPTM strategy that leverages real-time information, such as road conditions and traffic flow speed, to dynamically manage battery temperature and vehicle speed profiles, ensuring optimal performance under varying conditions. The main contributions are summarized as follows:
\begin{itemize}
    \item A novel strategy for the IPTM of CAEVs is proposed. This strategy consolidates multi-objective optimization into a single term, eliminating the need for cumbersome parameter weighting, and simultaneously ensures safety, traffic efficiency, energy efficiency, optimal temperature, and minimized battery degradation.
    \item By incorporating a battery pack model, the proposed approach accurately captures the coupled electrical, thermal, and aging dynamics, addressing battery degradation and its inconsistencies within the battery pack.
    \item Comprehensive simulation results validate the proposed strategy, demonstrating significant improvements in energy efficiency, battery longevity, and reduced degradation inconsistency compared to the benchmark strategy.
    \item The optimal selection of prediction horizons and sampling times for IPTM strategy in CAEVs is derived through rigorous verification and sensitivity analysis, ensuring robust performance across various operational scenarios. 
    \item We extract some novel insights to guide the design for IPTM strategy in addressing multiple trade-offs, i.e., between battery degradation and energy efficiency, by focusing on minimizing peak power and cooling the battery during periods of low vehicle traction power demand.
\end{itemize}

\section{System configuration and modeling}\label{model}
In this section, we present the configuration and modeling of the power and thermal management system for CAEVs. This includes an overview of the vehicle's longitudinal dynamics, the energy consumption model, and a comprehensive battery pack model encompassing electrical, thermal, and aging behaviors, along with the associated BTMS.

\subsection{Vehicle Longitudinal Dynamics}
The vehicle longitudinal dynamics can be described by its position $p$, velocity $v$ and acceleration $a$, shown as follows,
\begin{equation}
    \centering
    \dot{p} = v, \quad \dot{v} = a
\end{equation}
Then, the corresponding vehicle traction force $F_{\mathrm{v}}$ can be modeled by,
\begin{equation}
    \centering
    \begin{aligned}
        F_{\mathrm{v}} = mg\sin \theta + mgf\cos\theta+ \frac{1}{2}C_{\mathrm{D}}A\rho v^{2} + m\delta a 
    \end{aligned}
\end{equation}
where $m$ is the vehicle mass, $g$ is the acceleration due to gravity, $\theta$ is the road slope, $f$ is the rolling resistance coefficient, $C_{\mathrm{D}}$ is the aerodynamic drag coefficient, $A$ is the frontal area, $\rho$ is the air density, and $\delta$ is the vehicle rotational inertia coefficient.

Assuming the wheels do not slip, the electric motor torque $T_{\mathrm{m}}$ can be expressed as \cite{liang2024adaptive},
\begin{equation}
    \centering
    T_{\mathrm{m}} = 
    \begin{aligned}
        & \frac{F_{\mathrm{v}}r_{\mathrm{w}}}{i_{\mathrm{g}}i_{\mathrm{0}}\eta_{\mathrm{t}}^{\mathrm{sign(F_{\mathrm{v}})}}} 
    \end{aligned}
\end{equation}
where $i_\mathrm{g}$ and $i_\mathrm{0}$ are the transmission and final drive ratios, respectively. $r_\mathrm{{w}}$ denotes the vehicle tire radius, $\eta_{\mathrm{t}}$ denotes the drive-line efficiency, and $\mathrm{sign(\cdot)}$ is the signum function.

\subsection{Energy Consumption Model}
In this study, regenerative braking energy is considered as external input power. The total output power demand includes the traction power $P_{\mathrm{tra}}$, compressor power of the air-conditioning (AC) system $P_{\mathrm{cp}}$, and auxiliary power sources $P_{\mathrm{aux}}$, i.e., pump and fan in the BTMS. Note that $P_{\mathrm{aux}}$ is treated as constant for simplification \cite{wu2024optimal,dong2024predictive}. Thus, the total power at the battery pack terminal can be expressed as follows:
\begin{equation}
    \centering
    P_{\mathrm{b}} = \left\{
    \begin{aligned}
        & \frac{1}{\eta_{\mathrm{b}}} \left(P_{\mathrm{tra}} + P_{\mathrm{cp}} + P_{\mathrm{aux}}\right) && P_{\mathrm{tra}} \geq 0\\
        & \eta_{\mathrm{b}} \zeta P_{\mathrm{tra}} + \frac{1}{\eta_{\mathrm{b}}} \left(P_{\mathrm{cp}} + P_{\mathrm{aux}}\right) && P_{\mathrm{tra}} < 0
    \end{aligned}
    \right.
    \label{eq_power_tot}
\end{equation}
where $\eta_{\mathrm{b}}$ denotes the battery efficiency, $\zeta$ denotes the regenerative braking efficiency.

Then, the traction power can be calculated based on the approximated closed-form equation \cite{dong2022predictive}, 
\begin{equation}
    \centering
    P_{\mathrm{tra}} = \epsilon v T_{\mathrm{m}} + \sigma T_{\mathrm{m}}^{2}
\end{equation}
where $\epsilon$ and $\sigma$ are the empirical coefficients, i.e., $\epsilon = \frac{i_{\mathrm{g}}i_{\mathrm{0}}}{r_{\mathrm{w}}}$.

\subsection{Battery Pack Electrical-thermal-aging Model}
The overall dynamics and inconsistencies of a battery pack in electric vehicles (EVs) are influenced by two primary factors. First is the initial inconsistency determined by the manufacturer or previous utilization, which can be exacerbated by the coupled electrical, thermal, and aging dynamics of individual battery cells. Second is the battery pack's configuration, which includes the series and parallel arrangement of cells and the cooling structure. Building upon the existing work \cite{song2022progression}, we develop a control-oriented model to capture the coupled dynamics in the battery pack, including a pack-level and reduced-order thermal model. 

\begin{figure}[!htpb]
    \centering
    \includegraphics[width=0.45\textwidth]{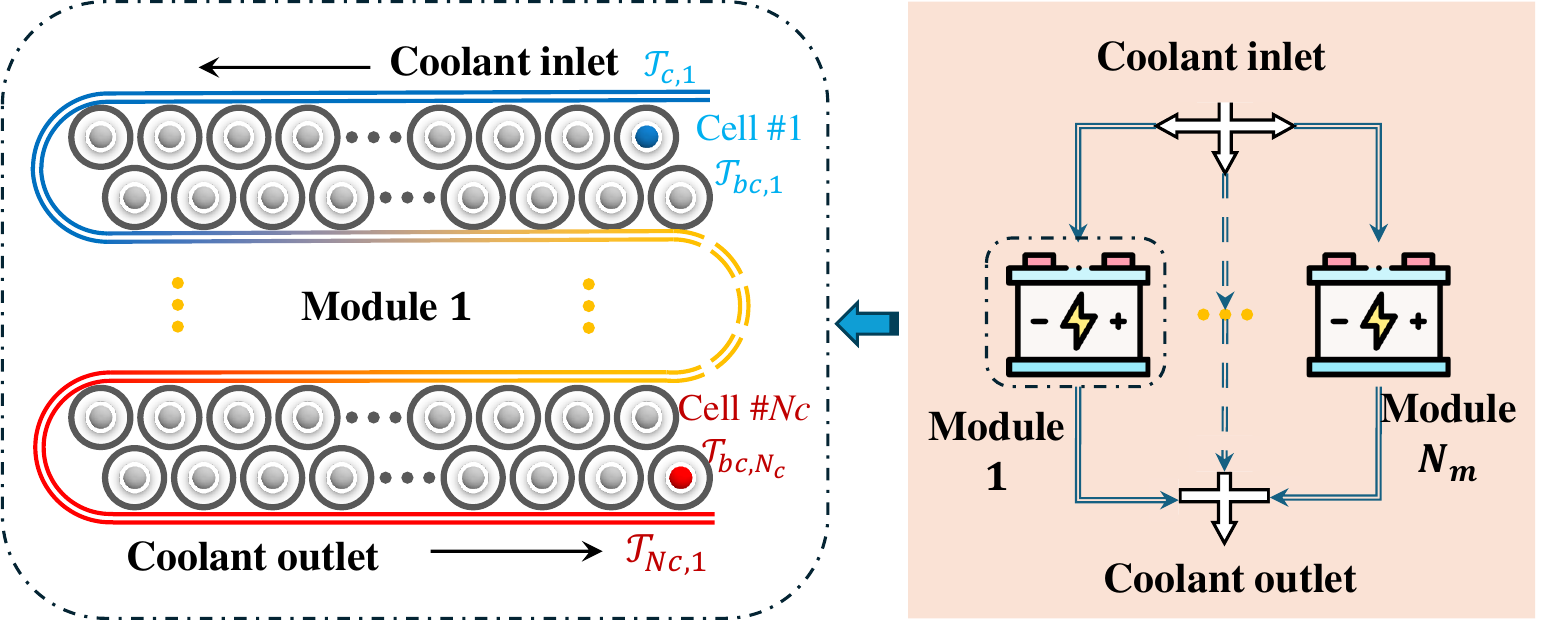}
    \caption{Schematic diagram of the cooling structure.}
    \label{fig:Coolingstructure}
\end{figure}

The battery pack is assumed to consist of $N_{\mathrm{m}}$ modules, with each module containing battery cells arranged in $N_{\mathrm{ms}}$ series and $N_{\mathrm{mp}}$ parallel configurations. Each module features an S-shaped cooling channel, which dissipates the heat generated by cells through direct contact, as illustrated in Fig. \ref{fig:Coolingstructure}. The inlet and outlet of each channel are connected in parallel across the modules. For simplification, we assume that all modules exhibit identical electrical, thermal, and aging dynamics.

\subsubsection{Electrical Model}
We employ an equivalent circuit model to capture the electrical dynamics of the battery. Specifically, the instantaneous voltage $V_{\mathrm{bc,i}}$ and current $I_{\mathrm{bc,i}}$ for cell $i$ are expressed as follows:
\begin{equation}
    \centering
    V_{\mathrm{bc,i}} = E_{\mathrm{bc,i}} - I_{\mathrm{bc,i}}R_{\mathrm{bc,i}}
    \label{eq_electrical}
\end{equation}
where $E_{\mathrm{bc,i}}$ denotes the open circuit voltage (OCV), and $R_{\mathrm{bc,i}}$ denotes the internal resistance. 

Assuming uniform current distribution across all cells, the variation in cell current is neglected. The battery pack's voltage $V_{\mathrm{b}}$, current $I_{\mathrm{b}}$ and power $P_{\mathrm{b}}$ can be represented as:
\begin{equation}
    \centering
    \left\{
    \begin{aligned}
        V_{\mathrm{b}} & = V_{\mathrm{bc,i}}\times N_{\mathrm{m}} \times N_{\mathrm{ms}} \\
        I_{\mathrm{b}} & = I_{\mathrm{bc,i}}\times N_{\mathrm{mp}} \\
        P_{\mathrm{b}} & = E_{\mathrm{b}}I_{\mathrm{b}} - \left( I_{\mathrm{b}}^{2}R_{\mathrm{b}} + I_{\mathrm{b}}T_{\mathrm{b}}\frac{dV_{\mathrm{b}}}{dT_{\mathrm{b}}} \right)
    \end{aligned}
    \right.
\end{equation}
where $E_{\mathrm{b}}$ denotes the battery pack OCV, $R_{\mathrm{b}}$ denotes the equivalent internal resistance of the battery pack, and $T_{\mathrm{b}}$ denotes the battery pack temperature.
\subsubsection{Thermal Model}
In the thermal model, we assume that heat transfer between cells can be neglected \cite{wei2024modeling}. Consequently, following our previous studies \cite{song2022progression,dong2024predictive}, the battery temperature is characterized by the cell surface temperature, which accounts for both heat generation from the current and heat dissipation through the coolant, as described below.
\begin{equation}
    \centering
    \dot{T}_{\mathrm{bc,i}} = \frac{I^{2}_{\mathrm{bc,i}}R_{\mathrm{bc,i}} + I_{\mathrm{bc,i}}T_{\mathrm{bc,i}}\frac{dV_{\mathrm{bc,i}}}{dT_{\mathrm{bc,i}}} + h\left( T_{\mathrm{c,i}} - T_{\mathrm{bc,i}} \right)}{C_{\mathrm{c}}}
    \label{eq_thermal}
\end{equation}
where $C_{\mathrm{c}}$ denotes the battery cell thermal capacity, $h$ denotes the convective heat transfer coefficient between the cell surface and the coolant, and $T_{\mathrm{c,i}}$ denotes the inlet coolant temperature at the $i^{\mathrm{th}}$ cell, which can be calculated as follows \cite{lin2019robust}:
\begin{equation}
    \centering
    T_{\mathrm{c,i}} = \left( 1-\frac{h}{C_{\mathrm{f}}} \right)^{\mathrm{i-1}}T_{\mathrm{c,in}} + \sum_{\mathrm{j}=2}^{\mathrm{i}}\frac{h}{C_{\mathrm{f}}}\left( 1 - \frac{h}{C_{\mathrm{f}}} \right)^{\mathrm{i-j}} T_{\mathrm{bc,j-1}}
    \label{eq_coolant}
\end{equation}
where $T_{\mathrm{c,in}}$ is the coolant temperature at the inlet of the coolant channel, which is also the inlet coolant temperature for the first cell $(T_{\mathrm{c,in}} = T_{\mathrm{c,1}})$, and $C_{\mathrm{f}}$ represents the thermal capacity of the coolant flow within a single channel, i.e.,
\begin{equation}
    \centering
    C_{\mathrm{f}} = C_{\mathrm{p}} \frac{\dot{m}}{N_{\mathrm{m}}} \Delta t
\end{equation}
where $\dot{m}$ represents the mass flow rate of the coolant, which is driven by the pump and assumed constant for simplification. $C_{\mathrm{p}}$ is the specific heat capacity of the coolant.

\subsubsection{Aging Model}
The degradation process of lithium-ion battery cells is described by a semi-empirical aging model \cite{wang2011cycle}, shown as follows,
\begin{equation}
    \centering
    Q_{\mathrm{loss}} = \frac{Q_{\mathrm{nom}} - Q_{\mathrm{bc}}}{Q_{\mathrm{nom}}} = Ae^{\left( \frac{-E_{\mathrm{a}} + B \cdot C_{\mathrm{Rate}}}{RT_{bc}} \right) }\left( A_{\mathrm{h}} \right)^{z}
    \label{eq_aging_contin}
\end{equation}
where $Q_{\mathrm{loss}}$ denotes the battery degradation, $Q_{\mathrm{nom}}$ denotes the cell nominal capacity, $Q_{\mathrm{bc}}$ denotes the cell remaining capacity at current cycle.  $E_{\mathrm{a}}$ denotes the activation energy, $A$ denotes the pre-exponential factor, $R$ denotes the gas constant, $A_{\mathrm{h}}$ denotes the amp-hour-throughput, $z$ denotes the exponential factor, $C_{\mathrm{Rate}}$ denotes the discharge (or charge) rate, and $B$ is the compensation factor. 

By discretizing the equation, a dynamic aging model can be developed as follows,
\begin{subequations}
    \begin{align}
        Q_{\mathrm{loss,k}} & = Q_{\mathrm{loss,k-1}} + \Delta Q_{\mathrm{loss,k}} \\
        \Delta Q_{\mathrm{loss,k}} & = \Delta A_{\mathrm{h}}zA^{\frac{1}{z}}e^{\frac{- E_{\mathrm{a}} + B\cdot C_{\mathrm{Rate}}}{zRT_{\mathrm{bc}}}}Q_{\mathrm{loss,k-1}}^{1-\frac{1}{z}} \label{eq_aging2_dis} \\
        \Delta A_{\mathrm{h}} & = \frac{1}{3600} \int_{t}^{t + \Delta t} \vert I_{\mathrm{bc}} \vert dt \\
        C_{\mathrm{Rate}} & = \frac{I_{\mathrm{bc}}}{Q_{\mathrm{nom}}}
    \end{align}
\end{subequations}

\subsubsection{Coupled Dynamics}
The coupled electrical-thermal-aging dynamics are interlinked through the internal resistance, $R_{bc}$. Initially, the sequential cooling structure results in temperature variations between battery cells (Eq. (\ref{eq_coolant})), which contributes to variations of internal resistance $R_{\mathrm{bc}}$. Then, the changing internal resistance subsequently affects heat generation in the thermal model (Eq. (\ref{eq_thermal})), leading to alterations in cell temperature. Consequently, the modified temperature impacts battery degradation in the aging model (Eq. (\ref{eq_aging_contin})). In turn, the degradation and temperature changes drive changes in the internal resistance, creating a close loop that forms a coupled dynamic system. Therefore, the effects of temperature and degradation on the internal resistance are described as follows \cite{song2022progression}:
\begin{equation}
    \centering
    \left\{
    \begin{aligned}
        R_{\mathrm{bc,i}} & = R_{\mathrm{bc,0}} + \kappa \left( T_{\mathrm{bc,i}} - T_{\mathrm{bc,0}} \right)R_{\mathrm{bc,0}} \\
        R_{\mathrm{bc,i}} & = \mu \left( \frac{Q_{\mathrm{non}}}{Q_{\mathrm{bc,j}}} \right)^{\lambda}R_{\mathrm{bc,0}}
    \end{aligned}
    \right.
\end{equation}
where $\kappa,\mu$ and $\lambda$ are pre-defined and constant coefficients $(\mu >1, \lambda \geq 1)$, $T_{\mathrm{bc,0}}$ denotes the nominal temperature (i.e., $15 ^\circ \mathrm{C}$), and $R_{\mathrm{bc,0}}$ denotes the internal resistance at nominal temperature and initial capacity \cite{song2014multi}.

By integrating the coupled dynamics of the battery pack, the battery pack electrical-thermal-aging model can be constructed. However, as the number of battery cells in a module increases, the complexity of the thermal dynamics escalates, resulting in a high computational burden that makes the model impractical for real-world applications. To strike a balance between computational efficiency and accuracy, we choose to model only the first and last cells, representing the minimum and maximum temperatures and degradation within the battery pack. Consequently, a simplified battery model capable of characterizing temperature and degradation inconsistencies is constructed.


\subsection{Battery Thermal Management System}
\begin{figure}[!htpb]
    \centering
    \includegraphics[width=0.45\textwidth]{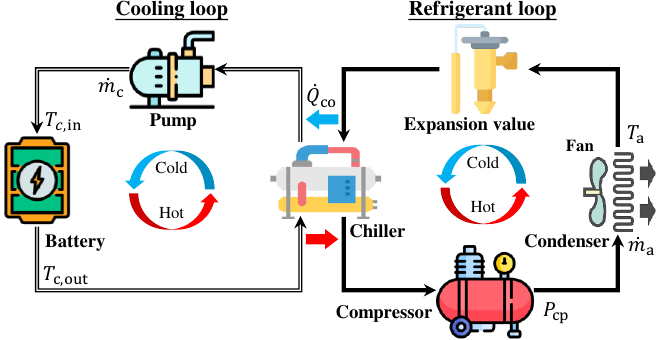}
    \caption{Schematic diagram of the studied BTMS.}
    \label{fig:BTMS}
\end{figure}

Fig. \ref{fig:BTMS} illustrates the schematic of the TMS under study. The system comprises two primary loops: the refrigeration loop and the battery cooling loop. In the refrigeration loop, the refrigerant is circulated by the compressor, releasing heat to the ambient air through the condenser and absorbing heat from the battery cooling loop via the chiller. Simultaneously, coolant in the battery cooling loop, driven by a pump, absorbs heat generated by the battery and transfers it to the refrigerant loop through the chiller.

Building on our previous study \cite{wu2024optimal}, a data-driven model is used to capture the complex, nonlinear dynamics of the BTMS. The relationship between the cooling rate $\dot{Q}_{\mathrm{co}}$ and the compressor power is derived from orthogonal simulation experiments, as detailed below. These experiments were conducted using the AC model developed in KULI software \cite{kiran2024case}.
\begin{equation}
    \centering
    \dot{Q}_{\mathrm{co}} = \xi_{1}P_{\mathrm{cp}} + \xi_{2}P_{\mathrm{cp}}^{2} + \xi_{3}T_{\mathrm{c,out}} + \xi_{4}T_{\mathrm{a}}\dot{m}_{\mathrm{a}} + \xi_{5}T_{\mathrm{c,out}}\dot{m}_{\mathrm{c}} + \xi_{6}
\end{equation}
where $T_{\mathrm{a}}$ denotes the ambient temperature, $\dot{m}_{\mathrm{a}}$ denotes the mass flow rate of air through the condenser and fan, characterized by $\dot{m}_{\mathrm{a}} = 0.07065 + 0.001683\cdot v$. $\xi$ denotes the fitted coefficients, determined based on $T_{\mathrm{a}}$ and $\dot{m}_{\mathrm{c}}$ \cite{wu2024optimal}. $T_{\mathrm{c,out}}$ denotes the outlet coolant temperature and can be considered as the inlet temperature at the $N_{\mathrm{c}} + 1$ cell. Hence, $T_{\mathrm{c,out}}$ and $T_{\mathrm{c,in}}$ can be calculated as follows,
\begin{equation}
    \centering
    \left\{
    \begin{aligned}
        T_{\mathrm{c,out}} & = \left( 1-\frac{h}{C_{\mathrm{f}}} \right)^{N_{\mathrm{c}}}T_{\mathrm{c,in}} + \sum_{\mathrm{j}=2}^{N_{\mathrm{c}}+1} T_{\mathrm{bc,j-1}} \\
        T_{\mathrm{c,in}}  & = T_{\mathrm{c,out}} - \frac{\dot{Q}_{\mathrm{co}}}{\dot{m}_{\mathrm{c}} C_{\mathrm{p}}}
    \end{aligned}
    \right.
\end{equation}

\section{Methodologies}
This section introduces the problem formulation, followed by an analysis of conventional control strategies and the proposed strategy.

\subsection{Problem Formulation}
\begin{figure}[!htpb]
    \centering
    \includegraphics[width=0.45\textwidth]{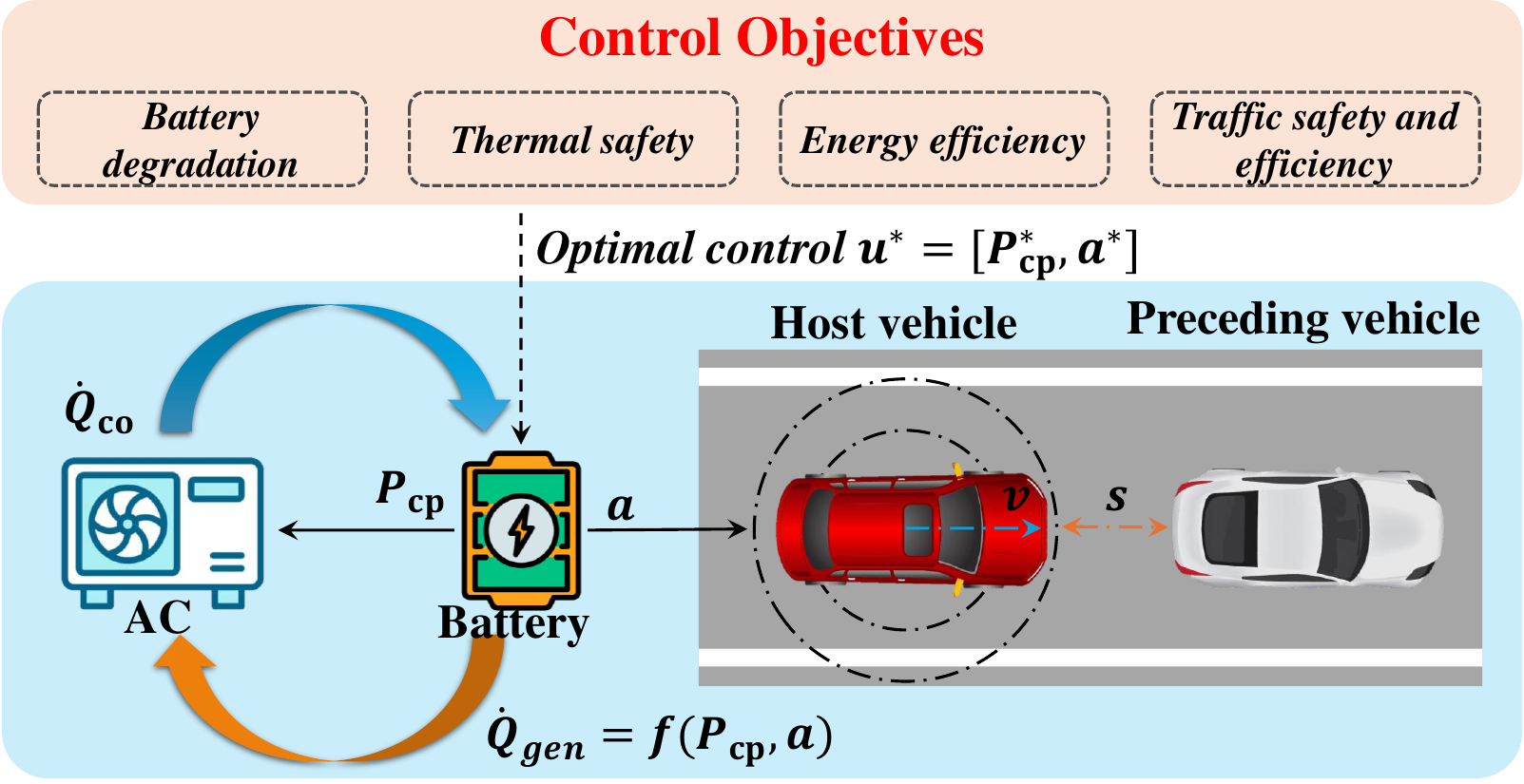}
    \caption{Schematic diagram of the IPTM system.}
    \label{fig:schematic}
\end{figure}

Fig. \ref{fig:schematic} illustrates the schematic diagram of the IPTM system. This configuration controls the host vehicle’s velocity to maintain safe traffic and energy-efficient driving while following the preceding vehicle. Concurrently, the battery pack's temperature is regulated to ensure thermal safety and minimize degradation. 

The key challenges of IPTM stem from two main factors: multi-timescale dynamics (i.e., the fast dynamics of the velocity and the slow dynamics of the temperature) and the complexities of multi-objective optimization with numerous trade-offs. 

Regarding multi-timescale dynamics, previous studies \cite{dong2024predictive,wu2024optimal,hu2021multihorizon,amini2020hierarchical} have demonstrated that the thermal response of the battery is significantly slower than the vehicle longitudinal dynamics. Conventional battery thermal management strategies, which rely solely on instantaneous temperature feedback to regulate cooling, often result in delayed cooling, particularly during periods of peak heat generation. This delay in response can lead to insufficient cooling during high-power demand scenarios, raising thermal safety concerns and increasing the energy required for battery cooling. 

In multi-objective optimization, balancing multiple objectives with complex trade-offs poses a significant challenge in generating optimal control actions. Specifically, in this IPTM system, three primary trade-offs must be considered: i) minimizing cooling energy consumption and reducing battery degradation, and ii) improving traction energy efficiency and traffic efficiency, and iii) maintaining optimal battery temperature while reducing energy consumption. For example, reducing battery degradation requires lowering temperature and energy consumption, but achieving lower temperatures necessitates higher cooling energy usage. These intricate interdependencies make it difficult to optimize all objectives simultaneously. 

To address these challenges, we propose an IPTM strategy, with the goal of ensuring safe, traffic-efficient, and energy-efficient driving while actively mitigating battery degradation through precise control of vehicle velocity and cooling power. 


Adopting the MPC framework, we define the state variable $x = [v, p, T_{\mathrm{bc,1}},T_{\mathrm{bc,N_{c}}},Q_{\mathrm{loss,1}},Q_{\mathrm{loss,N_{c}}}]$, which includes the vehicle's velocity, position, and the temperature and degradation of cell $1$ and cell $N_{\mathrm{c}}$.
And the control input $u = [a,P_{\mathrm{cp}}]$ includes the vehicle's acceleration and the compressor power for cooling. Meanwhile, the mass flow rate of coolant $\dot{m}$ and ambient temperature $T_{\mathrm{air}}$ are assumed as constants for simplification \cite{wu2024optimal}. The objective function $J$ is designed to minimize traffic safety risks, improve efficiency, regulate temperature, reduce energy consumption, and mitigate battery degradation. Additionally, inequality constraints are imposed on inter-vehicle spacing, velocity, temperature, acceleration, and its variations, as well as compressor power and its fluctuations. Hence, the optimization control problem is expressed as follows:
\begin{subequations}
    \begin{align}
        \underset{u\in\mathcal{U}}{\min} \quad & J(x,u) \label{eq_objective} \\
        \mathrm{s.t.} \quad & a \in [a_{\mathrm{min}}, a_{\mathrm{max}}]  \label{eq_cons1}  \\
        & \Delta a \in [\Delta a _{\mathrm{min}},\Delta a _{\mathrm{max}}]   \label{eq_cons2}  \\
        & P_{\mathrm{cp}} \in [P_{\mathrm{cp,min}},P_{\mathrm{cp,max}}]   \label{eq_cons3}  \\
        & \Delta P_{\mathrm{cp}} \in [\Delta P_{\mathrm{cp,min}}, \Delta P_{\mathrm{cp,max}}]   \label{eq_cons4}  \\
        & v \in [v_{\mathrm{min}},v_{\mathrm{max}}]   \label{eq_cons5}  \\
        & p_{\mathrm{PV}} - p \in [s_{\mathrm{min}},s_{\mathrm{max}}]   \label{eq_cons6}  \\
        & T_{\mathrm{bc,1}},T_{\mathrm{bc,N_{c}}} \in [T_{\mathrm{b,min}},T_{\mathrm{b,max}}]   \label{eq_cons7}  \\
        & T_{\mathrm{c,out}} \in [T_{\mathrm{c,out,min}},T_{\mathrm{c,out,max}}]   \label{eq_cons8}
    \end{align}
    \label{eq_cost_general}%
\end{subequations}
where the subscript $\mathrm{min}$ and $\mathrm{max}$ denote the lower and upper constraints, respectively. The values of these parameters are provided in Table \ref{tab1} \cite{wu2024optimal}. Additionally, we define $T_{\mathrm{b,min}} = T_{\mathrm{c,out,min}}$ and $T_{\mathrm{b,max}} = T_{\mathrm{c,out,max}}$. The terms $s_{\mathrm{min}}$ and $s_{\mathrm{max}}$ denote the safety and traffic efficiency constraints. Specifically, $s_{\min}$ can be calculated by the intelligent driver model \cite{treiber2000congested}, shown as follows:
\begin{equation}
    \centering
    \begin{aligned}
        s_{\mathrm{min}} & = s_{\mathrm{st}} + vT_{\mathrm{h}} + \frac{v(v-v_{PV})}{2\sqrt{-a_{\min}a_{\max}}} 
    \end{aligned}
\end{equation}
where $s_{\mathrm{st}}$ denotes the minimum inter-vehicle spacing, $T_{\mathrm{h}}$ denotes the safe time headway.

\begin{table}[htbp]
\caption{Parameters and values of constraints}
\begin{center}
\begin{tabular}{ll|ll}
\hline \hline
$a_{\mathrm{min}} $ & $\mathrm{-2}\mathrm{m/s^2}$ & $\Delta a_{\mathrm{min}} $  & $\mathrm{-0.5}\cdot \Delta t~ \mathrm{m/s^2}$  \\ 
\hline
$a_{\mathrm{max}} $      & $\mathrm{2}\mathrm{m/s^2}$       & $\Delta a_{\mathrm{max}}$ & $\mathrm{0.5}\cdot \Delta t~ \mathrm{m/s^2}$ \\
\hline \hline
$P_{\mathrm{cp,min}}$ & $\mathrm{0} \mathrm{W}$ & $\Delta P_{\mathrm{cp,min}}$ & $\mathrm{-200}\cdot \Delta t~ \mathrm{W}$  \\ 
\hline
$P_{\mathrm{cp,max}}$ & $\mathrm{4500} \mathrm{W}$  & $\Delta P_{\mathrm{cp,max}}$  & $\mathrm{200}\cdot \Delta t~ \mathrm{W}$ \\
\hline \hline
$v_{\mathrm{min}}$ & $\mathrm{0} \mathrm{km/h}$ & $T_{\mathrm{cout,min}}$ & $\mathrm{25}\mathrm{^\circ C}$  \\
\hline
$v_{\mathrm{max}}$ & $\mathrm{135}\mathrm{km/h}$& $T_{\mathrm{cout,max}}$ & $\mathrm{40}\mathrm{^\circ C}$\\  
\hline \hline
\end{tabular}
\label{tab1}
\end{center}
\end{table}

Regarding the objective function from Eq. (\ref{eq_objective}), the conventional strategy to achieve these goals is given as follows,
\begin{equation}
    \centering
    \begin{aligned}
        J(x,u) & = \sum_{k=0}^{N_{\mathrm{p}}-1} \left( J_{1} + J_{2} + J_{3} + J_{4} \right) \\
        J_{1}  & = \Vert v(k) - r_{\mathrm{v}} \Vert_{Q}^{2} \\
        J_{2}  & = \Vert T_{\mathrm{bc,1}}(k) - r_{\mathrm{T}} \Vert_{R}^{2} + \Vert T_{\mathrm{bc,N_{c}}}(k) - r_{\mathrm{T}} \Vert_{R}^{2} \\
        J_{3}  & = \lambda_{\mathrm{P}} \cdot P_{\mathrm{b}}(k) \cdot \Delta t \\
        J_{4}  & = \lambda_{\mathrm{Q}} \cdot \left( \Delta Q_{\mathrm{loss,1}}(k)  + \Delta Q_{\mathrm{loss,N_{c}}}(k) \right) 
    \end{aligned}
    \label{eq_cost_con}
\end{equation}
where $N_{\mathrm{p}}$ denotes the prediction horizon, $r_{\mathrm{v}}$ and $r_{\mathrm{T}}$ denote velocity and temperature reference, respectively. $J_{1}$ denotes the cost penalized by matrix $Q$ to ensure safe and efficient car-following behavior. $J_{2}$ denotes the cost penalized by matrix $R$ to maintain battery cells' operating temperature within a safe range. $J_{3}$ denotes the total energy consumption, penalized by $\lambda_{\mathrm{P}}$, aiming at minimizing energy usage. Finally, $J_{4}$ represents the sum of degradation loss of cell $1$ and $N_{\mathrm{c}}$, penalized by $\lambda_{\mathrm{Q}}$.

\subsection{Analysis of the Conventional Strategies}
Before introducing the proposed strategy, it is essential to examine the objective function in Eq. (\ref{eq_cost_con}) to better understand the rationale behind the novel strategy and the underlying challenges, including multi-timescale dynamics and inherent trade-offs.

To simplify the analysis, we initially omit the battery degradation loss term $J_{4}$. The primary goals of energy efficiency and thermal management necessitate the inclusion of $J_{2}$ and $J_{3}$, which rely on carefully calibrated weighting parameters, $R$ and $\lambda_{\mathrm{P}}$. Poor parameter tuning can lead to sub-optimal power and thermal performance. Moreover, the energy consumption term, $J_{3}$, impacts traffic flow efficiency by reducing vehicle speed and increasing inter-vehicle spacing. Therefore, incorporating a speed reference in $J_{1}$, with an appropriate weighting parameter, $Q$, is a straightforward strategy to mitigate the negative effects. 

In this context, two key trade-offs emerge: 1) between traffic flow efficiency and energy efficiency, and 2) between optimal temperature regulation and energy efficiency. Additionally, effectively addressing these trade-offs requires carefully tuned weighting parameters and extended prediction horizons.

Finally, when battery degradation loss is considered, tuning the weighting parameters becomes increasingly complex due to the added dimensionality and intricate interdependencies. Specifically, this introduces a new trade-off between cooling energy consumption and battery degradation. Therefore, it is essential to develop a novel strategy that balances these objectives while simplifying the parameter settings, which is crucial for optimizing the power and thermal management of CAEVs.

\subsection{Integrated Power and Thermal Management Strategy}
We first introduce the proposed IPTM strategy, followed by a comprehensive analysis to substantiate the approach.

The proposed IPTM strategy is built upon an aging model and utilizes a Multi-Horizon Model Predictive Control (MH-MPC) framework. Specifically, the aging model primarily addresses the trade-offs of multi-objectives. To address another challenge of multi-timescale dynamics, an MH-MPC framework using two receding horizons with different sampling times is employed to reduce the computational footprint by properly selecting different sampling times over different prediction horizons. For further details on MH-MPC, please refer to \cite{hu2021multihorizon,hu2020integrated}. The optimization problem is formulated as follows:
\begin{equation}
    \centering
    \begin{aligned}
        \underset{u\in \mathcal{U}}{\min} \quad & 
        J(x,u) = \sum_{k=0}^{N_{\mathrm{p1}}} J_{4} + \sum_{k=N_{\mathrm{p1}}+1}^{N_{\mathrm{p2}}} J_{4} \\
        \mathrm{s.t.} \quad & (\ref{eq_cons1}),(\ref{eq_cons2}),(\ref{eq_cons3}),(\ref{eq_cons4}),(\ref{eq_cons5}),(\ref{eq_cons6}), \\
        & (\ref{eq_cons7}), \text{ and } (\ref{eq_cons8})
    \end{aligned}
    \label{eq_novel}
\end{equation}
where $N_{\mathrm{p1}}$ and $N_{\mathrm{p2}}$ represent the short-term and long-term prediction horizons, corresponding to the sampling times $\Delta t_{1}$ and $\Delta t_{2}$, respectively. Note that a key advantage of the proposed approach is that it eliminates the need for tuning weighting parameters.

Building on the analysis of conventional strategies, it is evident that the cost function in Eq. (\ref{eq_cost_con}) embodies manually defined objectives. In contrast, in Eq. (\ref{eq_novel}), we simplify the complexity of the cost function by focusing solely on the battery degradation loss, $\Delta Q_{\mathrm{loss}}$, in the cost function. Although other objectives are not explicitly included, they can still be effectively addressed, as revealed by the aging model.

\begin{figure}[!htpb]
    \centering
    \includegraphics[width=0.4\textwidth]{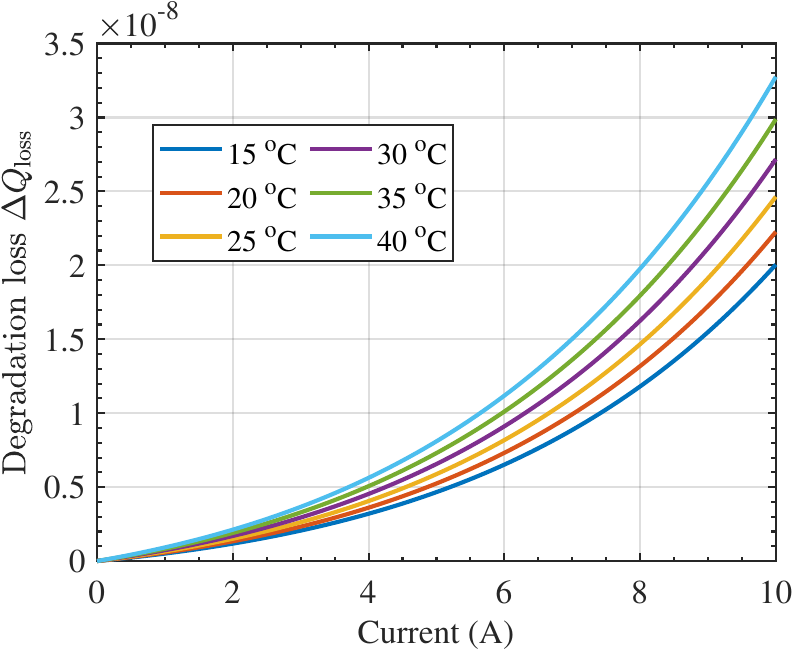}
    \caption{Influences of current and temperature on degradation ($10$A = $2$C).}
    \label{fig:aging}
\end{figure}

Fig. \ref{fig:aging} illustrates the battery degradation loss of a 5 Ah battery cell under varying currents and temperatures. This figure reveals that reducing the battery degradation loss can be achieved by either lowering the current or the operating temperature. Furthermore, since the battery power is directly proportional to the current (given the slow change in voltage, i.e., $P_{\mathrm{b}} = I_{\mathrm{b}} \cdot V_{\mathrm{b}}$), reducing the current throughout is effectively equivalent to reducing energy consumption.

From Eq. (\ref{eq_power_tot}), we know that the total current $I_{\mathrm{b}}$ is a combination of traction current $I_{\mathrm{tra}}$, compressor current $I_{\mathrm{cp}}$, and auxiliary current $I_{\mathrm{aux}}$. To develop an advanced power and thermal management strategy, two key insights emerge. i) \textbf{Exponential degradation sensitivity to current:} As shown in Fig. \ref{fig:aging}, battery degradation loss increases exponentially with current. ii) \textbf{Magnitude of currents:} The traction current is significantly larger than the compressor current and the auxiliary current, making it the dominant contributor to battery degradation.

Given these factors, an efficient strategy for power management can be formulated. During low-power demand periods, vehicle acceleration can be increased to enhance traffic efficiency (reducing inter-vehicle spacing), as the added degradation loss from low currents is minimal.  This stored inter-vehicle spacing can then be utilized during high-power demand periods, allowing reduced acceleration, thereby limiting peak current and minimizing degradation when the battery is most sensitive to high currents. Under such a strategy, While overall energy consumption may remain similar, the strategy effectively minimizes battery degradation.

From a thermal management perspective, it is essential to avoid activating the AC system for battery cooling during high traction power demand, as battery degradation is particularly sensitive under high currents. By reducing the total current, battery degradation can be minimized. Instead, cooling should be prioritized during low traction power periods because reduced battery temperatures contribute to lower degradation loss.
Additionally, such a strategy has a hidden effect. The heat generated by the compressor is proportional to the square of the total current, meaning that running the compressor during high traction demand increases heat generation, requiring more energy for cooling. By timing the cooling during low power periods, the battery temperature drops efficiently, minimizing battery degradation while also optimizing energy consumption.

Overall, by focusing exclusively on battery degradation loss in the cost function, the multi-objectives, i.e., regulating temperature, minimizing power, and mitigating battery degradation, can still be theoretically addressed to some extent.

\section{Results}

\subsection{Simulation Setup}
\begin{figure}[!htpb]
    \centering
    \includegraphics[width=0.48\textwidth]{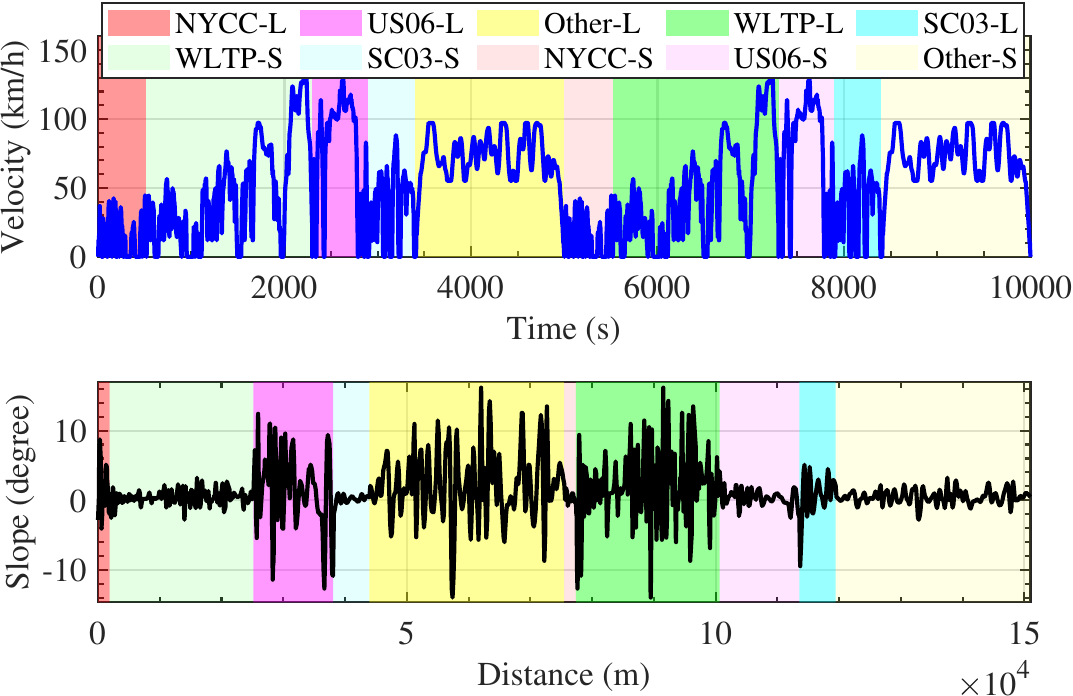}
    \caption{Comprehensive driving cycle. $\mathrm{L}$ and $\mathrm{S}$ represent large road slope and small road slope conditions, respectively.}
    \label{fig:driving_cycle}
\end{figure}

In this study, the proposed strategy is evaluated in a scenario where the host vehicle follows a preceding vehicle using V2X technology, as shown in Fig. \ref{fig:schematic}. This scenario assumes access to future speed profiles of the preceding vehicle, along with road slope information, allowing for more precise control decisions \cite{do2021application}. To assess the strategy’s effectiveness, a comprehensive driving cycle spanning $10^{4}$ seconds with a sampling interval of $T_{\mathrm{s}} =1\mathrm{s}$ is designed. The driving cycle comprises several standard and representative cycles, including NYCC, WLTP, US06, SC03, and additional smooth driving cycles, ensuring a wide coverage of driving conditions. To evaluate the robustness, two distinct road slope conditions are incorporated. One with a relatively large slope and the other with a small slope. Each driving cycle segment is subjected to both two slope types, and the entire cycle is repeated twice, as shown in Fig. \ref{fig:driving_cycle}. 

In conventional strategies, the velocity reference is set as the average velocity of the PV over the past two seconds, while the temperature references for both cell $1$ and cell $N_{\mathrm{c}}$ are fixed at $r_{\mathrm{T}} = 26^o\mathrm{C}$. Additionally, the weighting parameters for reference tracking are predetermined, with $Q=0.5$ and $R=0.1$. For other performance-related weighting parameters, we define $\lambda_{\mathrm{P}} \in [10^{-5},10^{-3}]$, $\lambda_{Q} \in [10^{6},10^{10}]$, and the ratio $\frac{\lambda_{\mathrm{Q}}}{\lambda_{\mathrm{P}}} \in [10^{6},10^{14}]$, based on the magnitudes of power and battery degradation loss. The vehicle parameters, including those related to the vehicle body, powertrain, battery, and BTMS, are outlined in Table \ref{tab:specifications}. 

We focus on analyzing the following three key aspects:

1) Various power and thermal management strategies are compared within the Single-Horizon MPC (SH-MPC) framework to ensure a fair evaluation of the effectiveness of the proposed strategy. 

2) Sensitivity analyses are performed to further validate the results and assess the influence of key parameters on system performance. 

3) In the Multi-Horizon MPC (MH-MPC) framework, we focus on identifying the key parameters of the proposed strategy to reduce computation time for practical implementation. Additionally, a brief comparison between SH-MPC and MH-MPC is provided to highlight their differences. This assessment is crucial for optimizing the strategy's performance across different prediction horizons and sampling times. 

\begin{table}[!htpb]
    \centering
    \caption{Vehicle parameters and values}
    \begin{tblr}{
    cell{1-9}{2-3} = {c=1}{l},
    cell{2}{1} = {r=8}{l},
    cell{10}{1} = {r=5}{l},
    cell{15}{1} = {r=10}{l},
    cell{25}{1} = {r=4}{l},
    hline{1,29} = {-}{0.9pt},
    hline{2,10,15,25} = {-}{}, 
    }
         Category     & Parameter                                        & Value \\
         Body         & Mass $m$                                         & $1432\mathrm{kg}$ \\
                      & Air density $\rho$                               & $1.026\mathrm{kg/m^{3}}$ \\
                      & Air drag coefficient $C_{\mathrm{D}}$            & $0.3$ \\
                      & Front area $A$                                   & $2.22\mathrm{m^2}$ \\
                      & Gravity acceleration $g$                         & $9.8\mathrm{m/s^2}$ \\
                      & Rotational inertia $\delta$                      & $1.022$ \\
                      & Rolling resistance $f$                           & $0.015$ \\
                      & Tire radius $r_{\mathrm{w}}$                     & $0.28\mathrm{m}$ \\
         Powertrain   & Driveline efficiency $\eta_{\mathrm{t}}$         & $0.9$ \\
                      & Empirical coefficient $\sigma$                   & $0.873$ \\
                      & Final drive ratio $i_{\mathrm{0}}$               & $3.789$ \\
                      & Regenerative braking efficiency $\zeta$          & $0.3$ \\
                      & Transmission ratio $i_{\mathrm{g}}$              & $2.80$ \\
         Battery      & Cell model type                                  & $18650$ \\
                      & Cell thermal capacity $C_{\mathrm{c}}$           & $45\mathrm{J/K}$ \\
                      & Cell capacity $Q_{\mathrm{nom}}$                 & $5.019\mathrm{Ah}$ \\
                      & Cell initial degradation $Q_{\mathrm{loss,ini}}$ & $0.001$ \\
                      & Open-circuit voltage $E_{\mathrm{b}}$            & $380\mathrm{V}$ \\
                      & Structure $N_{\mathrm{m}}\times N_{\mathrm{ms}}\times N_{\mathrm{mp}}$ & $16\times 6 \times 38$ \\
                      & Activation energy $E_{\mathrm{a}}$               &  $15162 \mathrm{J}$\\
                      & Pre-exponential factor $A$                       & $0.0032$\\
                      & Exponential factor $z$                           & $0.824$  \\
                      & Compensation factor $B$                          & $1516$ \\
         BTMS         & Convective heat transfer coefficient $h$         & $0.4901\mathrm{W/K}$ \\
                      & Auxiliary power $P_{\mathrm{aux}}$               & $200\mathrm{W}$ \\
                      & Mass flow rate of coolant $\dot{m}_{\mathrm{c}}$ & $0.144\mathrm{kg/s}$ \\
                      & Specific thermal capacity of coolant $C_{\mathrm{p}}$ & $3330\mathrm{J/kg/} ^\circ \mathrm{C}$
    \end{tblr}
    \label{tab:specifications}
\end{table}

\subsection{Performance Comparisons under SH-MPC}
The primary objective of this section is to validate the performance of the proposed strategy. To ensure a fair comparison, the SH-MPC framework is used, eliminating the influence of sampling times and prediction horizons. Six different formulations are selected for comparison, all utilizing a prediction horizon of $N_{\mathrm{p}} = N_{\mathrm{p1}} = 15$ and a uniform sampling time of $\Delta t_{1}=1 \mathrm{s}$. The objective functions differ based on various combinations of cost terms, as outlined in Eq. (\ref{eq_cost_con}), as detailed in Table \ref{tab:formulation_comparison}. Among these, the reference tracking formulation serves as the primary benchmark. The conventional strategies include combinations of reference with energy $J_{3}$, aging $J_{4}$, or both. The strategy combining energy and aging costs serves as the vice benchmark, while the formulation focusing solely on aging represents the proposed approach.

\begin{table}[!htpb]
    \centering
    \caption{Formulation settings: objective function}
    \begin{tblr}{
    cell{1-4}{1-4} = {c=1}{c},
    cell{5}{1} = {c=4}{l},
    hline{1,3,5} = {-}{0.9pt},
    hline{2,4}   = {-}{},
    }
         Item          & Reference     & Energy+Aging & Aging     \\
         Objective     & $J_{1}+J_{2}$ & $J_{3}+J_{4}$  & $J_{4}$   \\
         Item          & Ref+Energy  & Ref+Aging    & Ref+Energy+Aging     \\
         Objective     & $J_{1}+J_{2}+J_{3}$ & $J_{1}+J_{2}+J_{4}$  & $J_{1}+J_{2}+J_{3}+J_{4}$   \\
         Note, the expression $J_{1}+J_{2}+J_{3}$ can be simplified to $J_{1-3}$.& & &
    \end{tblr}
    \label{tab:formulation_comparison}
\end{table}

\subsubsection{Comparison of Statistical Results}
In addition to the benchmark strategy and the proposed strategy, the performance of all other strategies relies on the tuning of the weighting parameters. To eliminate the influence of weighting parameters, the most promising results obtained from the trials of each strategy are selected based on the subsequent sensitivity analysis.

Table. \ref{tab:comparison_results} presents a detailed comparison of the outcomes, demonstrating that all revised formulations successfully reduce cooling energy, traction energy, total energy, and degradation loss in both cell $1$ and cell $N_{\mathrm{c}}$.

Starting with the conventional strategies, incorporating the energy term $J_{3}$ into the benchmark strategy leads to greater improvements across all metrics compared to incorporating the aging term $J_{4}$. This is primarily because energy consumption, which is proportional to current, is a major contributor to battery degradation. Therefore, reducing energy consumption directly extends the lifespan of lithium batteries. While the aging term $J_{4}$ not only indirectly optimizes energy consumption, but also increases energy demand for battery cooling compared to the energy term. The latter effect partially overlaps with the temperature reference $J_{2}$, leading to smaller reductions in cooling energy, traction energy, and associated battery degradation. Consequently, the objective function incorporating both energy $J_{3}$ and aging $J_{4}$ terms achieves intermediate performance improvements between the two individual strategies.

Turning to the other strategies, the vice benchmark strategy $J_{3,4}$, consistently outperforms all conventional strategies across nearly all metrics. This suggests that tracking speed and temperature references can significantly limit optimality, particularly when these references are sub-optimal. Furthermore, the proposed strategy $J_{4}$ further amplifies reductions in energy consumption, battery degradation loss, and degradation inconsistency compared to the vice benchmark. Specifically, while the vice benchmark achieves a $6.07\%$ reduction in energy, nearly a $5\%$ reduction in battery degradation loss, and over $9\%$ reduction in degradation inconsistency, the proposed strategy delivers even greater improvements: a $7.32\%$ reduction in energy, more than $12\%$ reduction in battery degradation loss, and a $30.36\%$ reduction in degradation inconsistency. As a result, the proposed strategy offers unmatched performance in power and thermal management for CAEVs.
\begin{table*}[!htpb]
	\centering
	\caption{Comparisons among different strategies.}
	\begin{tblr}{
	  cell{1}{2} = {c=6}{c},
        cell{2-15}{1-7} = {c=1}{c},
        cell{4}{2} = {r=2}{c},
        cell{6}{2} = {r=2}{c},
        cell{8}{2} = {r=2}{c},
        cell{10}{2} = {r=2}{c},
        cell{12}{2} = {r=2}{c},
        cell{14}{2} = {r=2}{c},
        cell{1}{1} = {r=3}{c},
        cell{16}{1} = {c=7}{l},
	  hline{1,4,16} = {-}{1pt},
	  hline{2} = {2-7}{},
        hline{6,8,10,12,14} = {1-7}{},
	}
		  & Strategies &    &    &   &   & \\
		  & Reference  & Ref + Energy & Ref + Aging & Ref + Energy + Aging  & Energy + Aging  & Aging  \\
            & $J_{1,2}$ & $J_{1-3}$  & $J_{1,2,4}$ & $J_{1-4}$ & $J_{3,4}$ & $J_{5}$ \\
	Cooling energy                  & $8049.01$   & $6925.07$    & $7903.57$    & $7549.34$    & $6165.48$    & $6492.42$    \\
        $\int \left(P_{\mathrm{cp}}+P_{\mathrm{aux}}\right)\cdot t~\mathrm{(kJ)}$   &              & $(-13.96\%)$ & $(-1.81\%)$  & $(-6.21\%)$  & $(-23.40\%)$ & $(-19.34\%)$  \\
	Traction energy                 & $179033.67$ & $172377.17$  & $178511.69$  & $175992.75$  & $169563.81$  & $166888.65$  \\
        $\int P_{\mathrm{tra}}\cdot t~\mathrm{(kJ)}$  &              & $(-3.72\%)$  & $(-0.29\%)$  & $(-1.70\%)$  & $(-5.29\%)$  & $(-6.78\%)$  \\
	Total energy                    & $187082.68$ & $179302.24$  & $186415.26$  & $183542.09$  & $175729.29$  & $173381.07$  \\
        $\int P_{\mathrm{b}}\cdot t~\mathrm{(kJ)}$    &              & $(-4.16\%)$  & $(-0.36\%)$  & $(-1.89\%)$  & $(-6.07\%)$  & $(-7.32\%)$  \\
	Cell $1$ degradation            & $1.2789$    & $1.2352$     & $1.2568$     & $1.2312$     & $1.2109$     & $1.1156$     \\
        $\Delta Q_{\mathrm{loss,1}}~(10^{-5})$ &             & $(-3.42\%)$  & $(-1.73\%)$  & $(-3.73\%)$ & $(-4.68\%)$  & $(-12.77\%)$ \\
	Cell $N_{c}$ degradation        & $1.3718$    & $1.3164$     & $1.3469$     & $1.3171$     & $1.3035$     & $1.1803$     \\
        $\Delta Q_{\mathrm{loss,N_c}}~(10^{-5})$ &             & $(-4.04\%)$  & $(-1.82\%)$  & $(-3.99\%)$ & $(-4.98\%)$  & $(-13.96\%)$ \\
        Degradation inconsistency       & $9.29$      & $8.12$       & $9.01$       & $8.59$       & $8.45$       & $6.47$       \\
        $Q_{\mathrm{loss,N_c}} - Q_{\mathrm{loss,1}}~(10^{-7})$      &              & $(-12.59\%)$ & $(-3.01\%)$  & $(-7.53\%)$ & $(-9.04\%)$  & $(-30.36\%)$ \\
        Note: Degradation inconsistency is quantified via the degradation difference between cell $N_{\mathrm{c}}$ and cell $1$. &  &  &  &  &  &
	\end{tblr}
	\label{tab:comparison_results}
\end{table*}
\subsubsection{Comparison of Control Trajectories}
\begin{figure}[!htb]
     \centering
     \subfigure[Total power distributions.]{
         \includegraphics[width=0.45\textwidth]{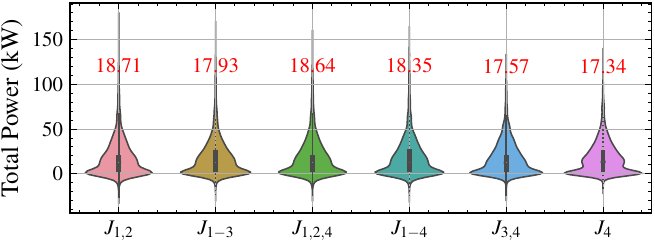}
         \label{fig_power_comparison}
     }
     \hfill
     \subfigure[Statistical power distributions.]{
         \includegraphics[width=0.45\textwidth]{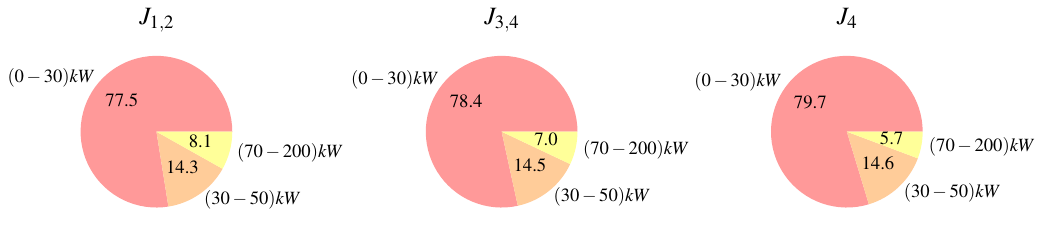}
         \label{fig_power_statistical}
     }
     \caption{Total power comparisons among different strategies.}
     \label{fig_power}
\end{figure}

\begin{figure}[!htb]
     \centering
     \subfigure[Temperature changes of Cell $1$ and $N_{\mathrm{c}}$.]{
         \includegraphics[width=0.45\textwidth]{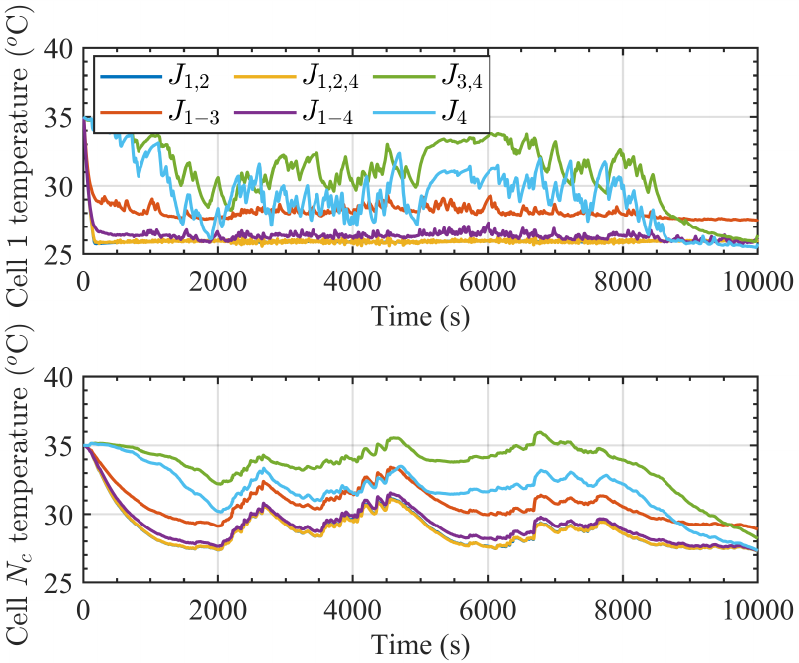}
         \label{fig_temperature_comparison}
     }
     \hfill
     \subfigure[Degradation changes of Cell $1$ and $N_{\mathrm{c}}$.]{
         \includegraphics[width=0.45\textwidth]{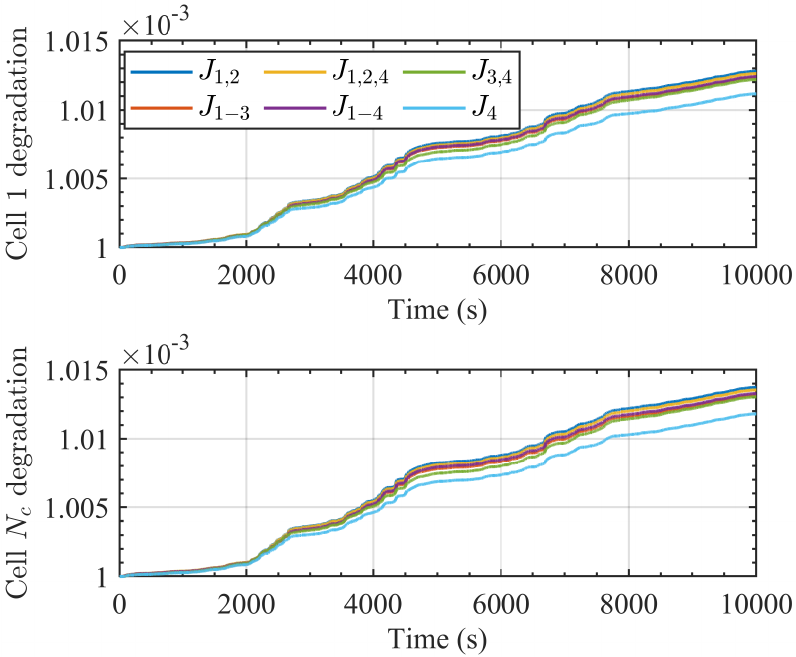}
         \label{fig_aging_comparison}
     }
     \caption{Temperature and degradation comparisons among different strategies.} 
     \label{fig_comparison_traj}
\end{figure}

Further analysis of the total power distribution, temperature profiles, and battery degradation profiles for all strategies is illustrated in Figs. \ref{fig_power} and \ref{fig_comparison_traj}, corresponding to the results in Table. \ref{tab:formulation_comparison}. Notably, the temperature changes shown in Fig. \ref{fig_temperature_comparison} highlight significant variations in cooling strategies and their impact on degradation outcomes.

Specifically, Fig. \ref{fig_power_comparison} illustrates the distribution and average total output power across different strategies, highlighting both the maximum and minimum power levels. Among these strategies, the proposed method achieves the lowest average total power, with a narrower distribution in high-power regions and a more concentrated presence in low-power areas.
In detail, the statistical total power distributions between the proposed strategy and two benchmark strategies, shown in Fig. \ref{fig_power_statistical}, clearly depict the low, medium, and high power regions. Such a power management strategy significantly contributes to battery degradation loss minimization. Once again, although the vice benchmark strategy achieves comparable energy consumption to the proposed method, there remains a significant difference in its ability to minimize battery degradation. This superior power management performance underscores the effectiveness of the proposed strategy.

From Fig. \ref{fig_comparison_traj}, the benchmark strategy consumes more energy to drive the AC system for battery cooling, resulting in a rapid temperature drop. However, due to the slower thermal dynamics of the last cell, $N_{\mathrm{c}}$, its temperature decreases more gradually compared to cell $1$. Additionally, under the benchmark strategy, both cells reach the lowest temperatures among all strategies, but this is accomplished by the highest degradation loss, as shown in Fig. \ref{fig_aging_comparison}. In contrast, incorporating energy consumption into the objective function, rather than focusing on aging alone, leads to greater reductions in degradation. This is because focusing on energy efficiency lowers overall energy use, which in turn reduces the current throughout the battery cells.

\begin{figure}[!htb]
     \centering
     \subfigure[Traction power and cooling power profiles.]{
         \includegraphics[width=0.45\textwidth]{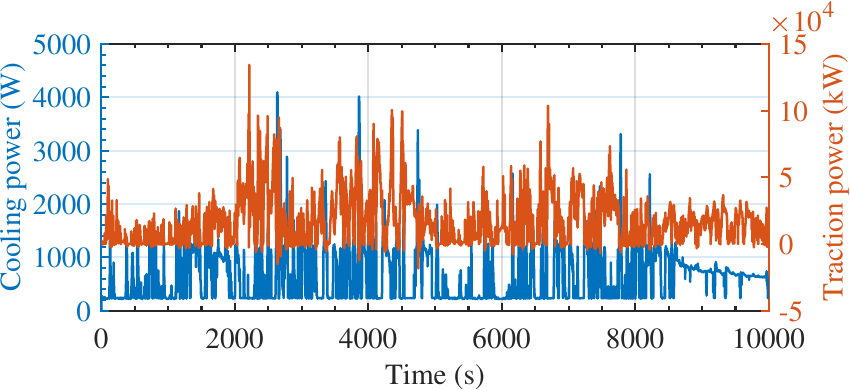}
         \label{fig_aging_cooling}
     }
     \hfill
     \subfigure[Inter-vehicle spacing (The minimum spacing remains larger than the safe threshold, i.e., $\min(p-p_{\mathrm{PV}}-s_{\min}) = 5.904\mathrm{m}$.]{
         \includegraphics[width=0.45\textwidth]{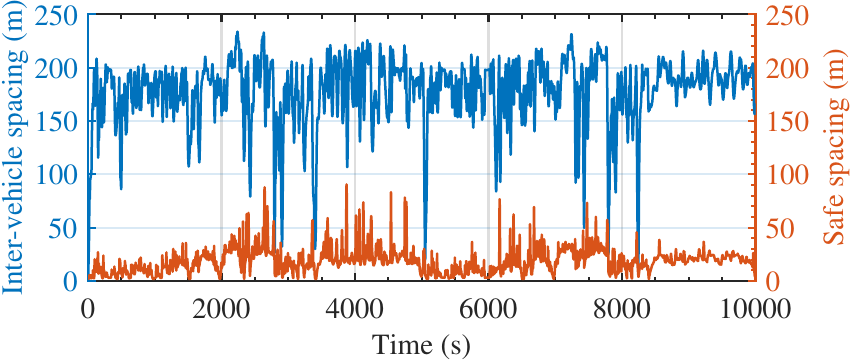}
         \label{fig_aging_spacing}
     }
     \caption{Traction power, cooling power, and inter-vehicle spacing profiles under the proposed strategy.}
     \label{fig_aging_power_spacing}
\end{figure}

The vice benchmark achieves the second-best overall performance, but its impact on reducing battery degradation is less pronounced than the proposed strategy. In contrast, the proposed strategy delivers significantly more battery degradation reduction than the other strategies, with both cells' temperatures lower than the vice benchmark, stabilizing around $30^o\mathrm{C}$. Meanwhile,  the proposed strategy consumes slightly more cooling energy than the vice benchmark, it achieves a notable reduction in temperature. This improvement is primarily due to the aging term in the proposed strategy, which effectively balances energy consumption and battery cooling. Specifically, it avoids consuming excessive energy use for cooling the battery under high traction power, while increasing cooling during low traction power or braking phases. This strategy substantially reduces the current load and external heat generation, ensuring temperature reduction, as seen in Fig. \ref{fig_aging_cooling}. Moreover, Fig. \ref{fig_aging_spacing} shows that the minimum spacing remains safely above the threshold, with at least $5.904\mathrm{m}$, confirming that the proposed strategy effectively maintains traffic safety under diverse driving conditions.

In summary, the proposed strategy effectively enhances energy efficiency, reduces battery degradation, and ensures consistency, all while ensuring traffic safety.

\subsection{Sensitivity Analysis}
In this section, two sensitivity analyses are conducted. The first focuses on the conventional strategies and the vice benchmark strategy to highlight the challenge of selecting suitable weighting parameters. The second demonstrates the dependency of the proposed strategy on the prediction horizon.

\subsubsection{Sensitivity Analysis on Weighting Parameters}
To validate the impact of the weighting parameters, we define four sets of values, ranging from small to large, for each strategy. The weighting parameters for reference tracking are fixed. Specifically, $\lambda_{\mathrm{P}} = [1\times 10^{-5}, 5\times 10^{-5}, 5\times 10^{-4}, 1\times 10^{-3}]$ for conventional strategy $J_{1-3}$, $\lambda_{\mathrm{Q}} = [1\times 10^6, 1\times 10^6, 5\times 10^7, 1\times 10^8]$ for conventional strategy $J_{1,2,4}$, and $\frac{\lambda_{\mathrm{Q}}}{\lambda_{\mathrm{P}}} = [10^{11}, 10^{12}, 10^{13}, 10^{14}]$ for conventional strategy $J_{1-4}$. For the vice benchmark strategy, we choose $\frac{\lambda_{\mathrm{Q}}}{\lambda_{\mathrm{P}}} = [5\times 10^6, 5\times 10^{10}, 5\times 10^{11}, 5\times 10^{12}]$. 

\begin{figure}[!htpb]
    \centering
    \includegraphics[width=0.45\textwidth]{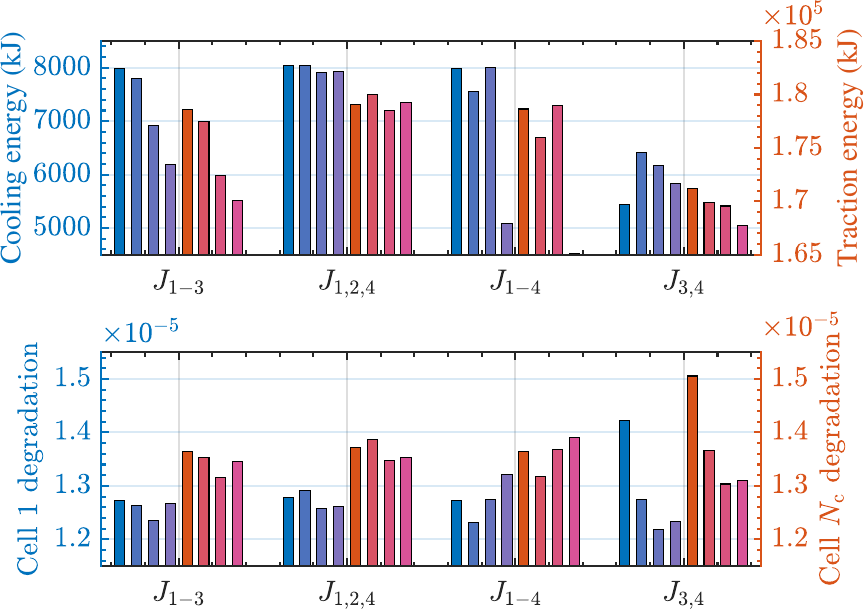}
    \caption{Sensitivity analysis on weighting parameters of conventional strategies.} 
    \label{fig:sensitivity_weighting}
\end{figure}

Fig. \ref{fig:sensitivity_weighting} illustrates the cooling energy, traction energy, and battery degradation of cell $1$ and cell $N_{\mathrm{c}}$. The blue series represents the cooling energy and battery degradation loss for cell $1$, while the red series represents the traction energy and battery degradation loss for cell $N_{\mathrm{c}}$. Variations in shades within each series indicate changes in weighting parameter values.

For the conventional strategy $J_{1-3}$, as the weighting parameter $\lambda_{\mathrm{P}}$ for energy consumption increases, both cooling and traction energy are significantly reduced. However, degradation loss does not always decrease accordingly, underscoring the trade-off between cooling energy consumption and battery degradation. Excessive reduction in cooling energy can lead to increased degradation. This trade-off is also evident in other strategies, especially in strategy $J_{1-4}$.
Regarding conventional strategy $J_{1,2,4}$, emphasizing the aging term does not result in substantial performance improvement. Finally, in the vice benchmark strategy $J_{4}$, as the ratio $\frac{\lambda_{\mathrm{Q}}}{\lambda_{\mathrm{P}}}$ increases, the traction energy consumption is significantly reduced. However, the battery degradation loss does not always follow the same trend. Prioritizing energy reduction through lower weighting ratios leads to decreased cooling energy consumption but at the cost of increased battery degradation loss.

In summary, the performance of the conventional strategies and the vice benchmark strategy is highly sensitive to the tuning of weighting parameters. As the number of objectives increases, tuning these parameters becomes increasingly complex. Moreover, the performance gains of these strategies are modest compared to the proposed strategy.

\subsubsection{Sensitivity Analysis on Prediction Horizon} 
\begin{figure}[!htpb]
    \centering
    \includegraphics[width=0.45\textwidth]{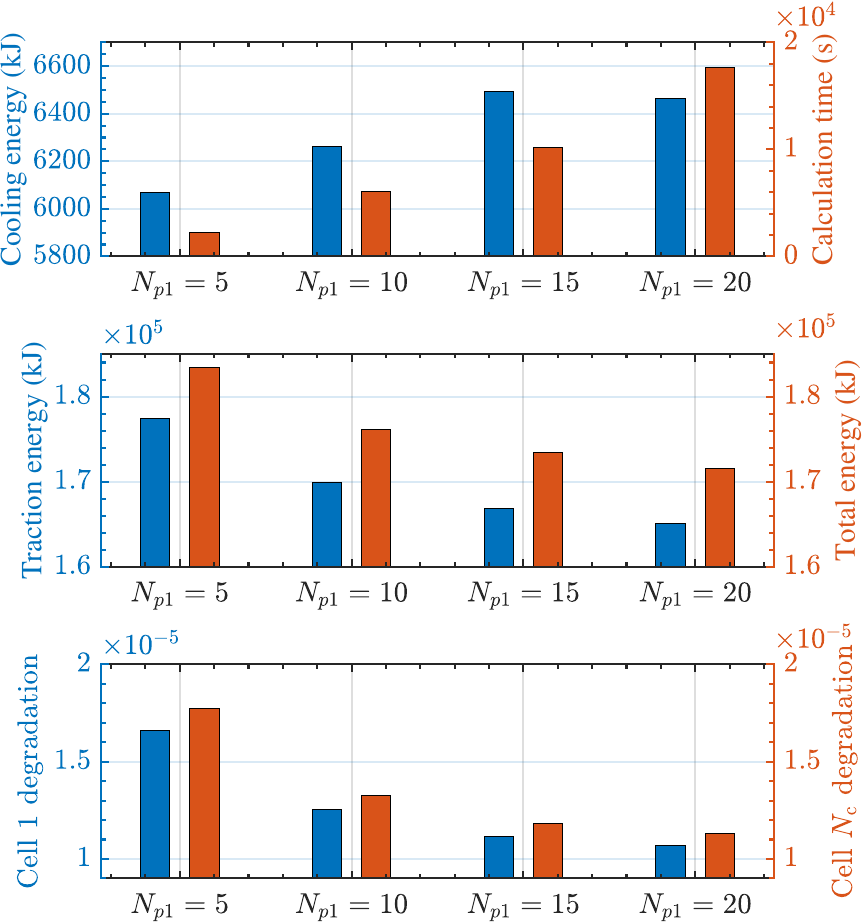}
    \caption{Sensitivity analysis on prediction horizon of the proposed strategy.}
    \label{fig:sensitivity_analysis_horizon}
\end{figure}

\begin{figure}
    \centering
    \includegraphics[width=0.45\textwidth]{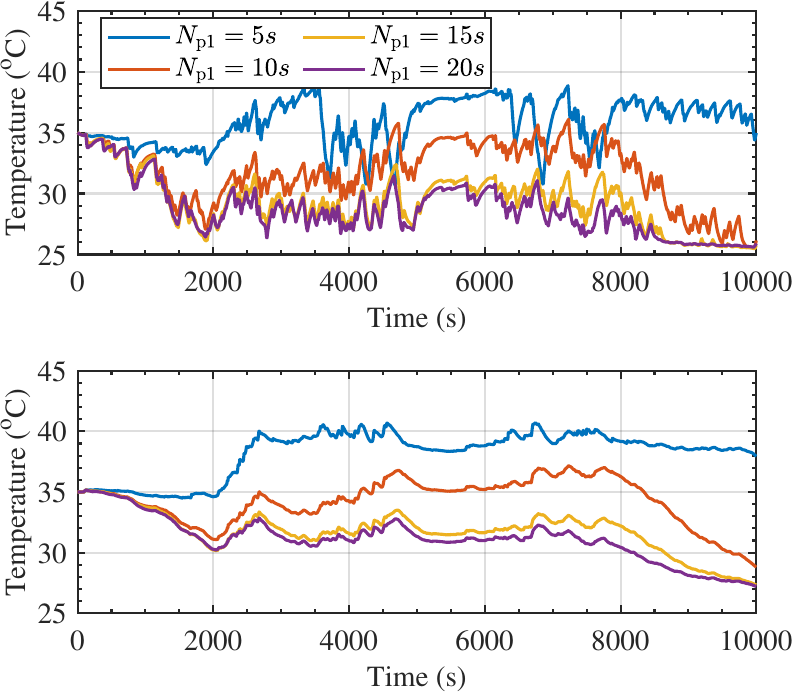}
    \caption{Temperature Changes of Cell $1$ and Cell $N_{\mathrm{c}}$ under the proposed strategy with different prediction horizons.}
    \label{fig:sensitivity_temperature}
\end{figure}

To assess the impact of the prediction horizon, we define four sets of values, ranging from $5$ seconds to $20$ seconds. The statistical results of the sensitivity analysis for various prediction horizons are summarized in Fig. \ref{fig:sensitivity_analysis_horizon},  with the corresponding temperature trajectories shown in Fig. \ref{fig:sensitivity_temperature}. 

From Fig. \ref{fig:sensitivity_analysis_horizon}, it can be observed that increasing the prediction horizon leads to a significant reduction in both traction energy consumption and battery degradation loss, although this comes at the expense of a considerably higher computational footprint. Additionally, the energy consumption for battery cooling tends to rise as the prediction horizon increases. This indicates that the proposed strategy is highly sensitive to the prediction horizon. A longer horizon allows for a more balanced trade-off between battery degradation and energy consumption, optimizing the timing of power demands to activate the AC system for battery cooling. Conversely, a shorter prediction horizon can negatively impact battery lifespan due to sub-optimal cooling strategies, as it tends to prioritize reducing power demand to minimize degradation while overlooking the impact of battery temperature. For instance, with a $5$ seconds prediction horizon, the strategy uses minimal energy for cooling, barely keeping the temperature of cell $N_{\mathrm{c}}$ within the constraints, as shown in Fig. \ref{fig:sensitivity_temperature}. This results in significantly higher degradation compared to the benchmark.

Additionally, during the period of high traction power demand between $2000\mathrm{s}$ and $2800\mathrm{s}$, caused by aggressive accelerations or road slopes, the temperatures of both cell $1$ and cell $N_{\mathrm{c}}$ increase substantially. This indicates that the proposed strategy avoids consuming energy for battery cooling during high-demand intervals. Furthermore, with a prediction horizon of at least $10$ seconds, the temperature is significantly reduced prior to $2000\mathrm{s}$, effectively preparing the system for the upcoming high traction demand. It is important to note that this strategy is primarily influenced by the aging term, as the prediction horizon is not long enough (i.e., hundreds of seconds) to capture future conditions. 
Fig. \ref{fig:sensitivity_temperature} shows that while the temperatures of both cell $1$ and $N_{\mathrm{c}}$ are lowest with a $N_{\mathrm{p1}}$ prediction horizon, the energy consumed to by the AC system for battery cooling is also lower compared to a $15\mathrm{s}$ horizon, as shown in Fig. \ref{fig:sensitivity_analysis_horizon}. This further suggests that a longer prediction horizon enables more efficient thermal management, particularly in accurately timing energy consumption for cooling the battery.

\subsection{Determining Suitable MH-MPC Controller} As previously mentioned, while the proposed strategy demonstrates promising performance, its computational poses challenges for practical application. To mitigate this issue, we reformulated the proposed strategy using the MH-MPC framework and evaluated its performance with different sampling times and prediction horizons. For simplicity, the sampling time for the short horizon is fixed at $\Delta t_{1} = 1\mathrm{s}$. Additionally, the total prediction horizon is set to $N_{\mathrm{p}} = N_{\mathrm{p1}} + N_{\mathrm{p2}} = 15$, with all prediction information assumed to be accurate.

\subsubsection{Sampling Time Determination for Long Horizon} 
\begin{figure}
    \centering
    \includegraphics[width=0.49\textwidth]{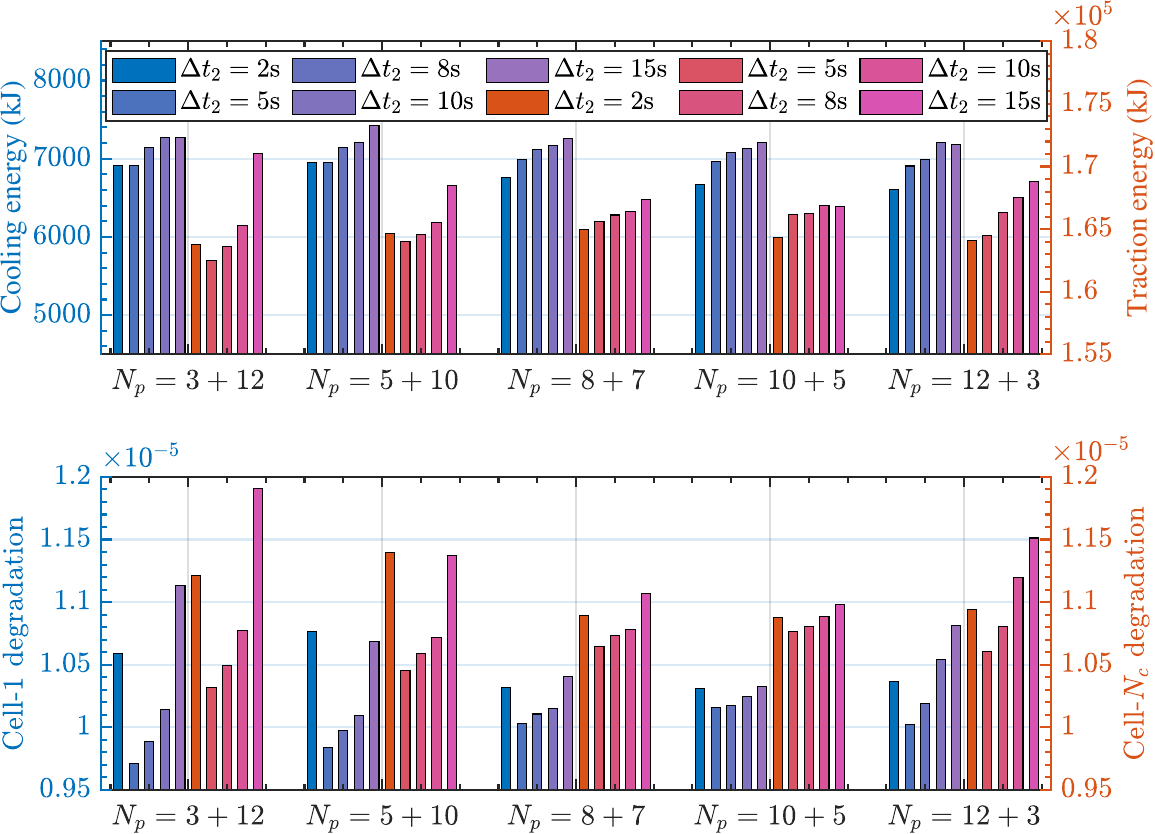}
    \caption{Sensitivity analysis on long horizon sampling times of the proposed strategy.}
    \label{fig:MH_MPC_step}
\end{figure}

We begin by determining the appropriate sampling time for the long horizon. Five candidate values are tested for the sampling times: $\Delta t_{2} = [2\mathrm{s}, 5\mathrm{s}, 8\mathrm{s}, 10\mathrm{s}, 15\mathrm{s}]$. Simultaneously, five combinations of short and long prediction steps are considered: $N_{\mathrm{p1}} + N_{\mathrm{p2}} = [3+12, 5+10, 8+7, 10+5, 12+3]$. The simulation results are summarized in Fig. \ref{fig:MH_MPC_step}.

From Fig. \ref{fig:MH_MPC_step}, it can be observed that the sampling time for the long horizon significantly impacts cooling energy, traction energy, and degradation. Specifically, increasing the sampling time leads to higher cooling energy consumption across all strategies. In most cases, traction energy consumption rises, contributing to higher battery degradation loss. These results highlight the importance of selecting an appropriate sampling time for the long horizon. Based on the findings in Fig. \ref{fig:MH_MPC_step}, the optimal sampling time for the long horizon is $\Delta t_{2} = 5\mathrm{s}$, as this set achieves the best degradation performance. Moreover, it is noteworthy that the MH-MPC strategy consistently results in lower battery degradation across all short-long prediction horizon combinations and sampling times compared to the SH-MPC strategy, highlighting the advantages of MH-MPC.

\subsubsection{Prediction Horizon Determination} From Fig. \ref{fig:MH_MPC_step}, it is evident that the optimal balance between degradation and energy efficiency is achieved with the set $N_{\mathrm{p1}} = 3$, while the second-best results are observed for $N_{\mathrm{p1}} = 5$. Consequently, these two sets, combined with the sampling times $\Delta t_{1} = 1\mathrm{s}$ and $\Delta t_{2} = 5\mathrm{s}$, are selected to assess the impact of varying prediction horizons. Specifically, five candidate values are tested for the long prediction horizon: $N_{\mathrm{p2}} = 5, 10, 12, 15, 20$.

\begin{figure}
    \centering
    \includegraphics[width=0.45\textwidth]{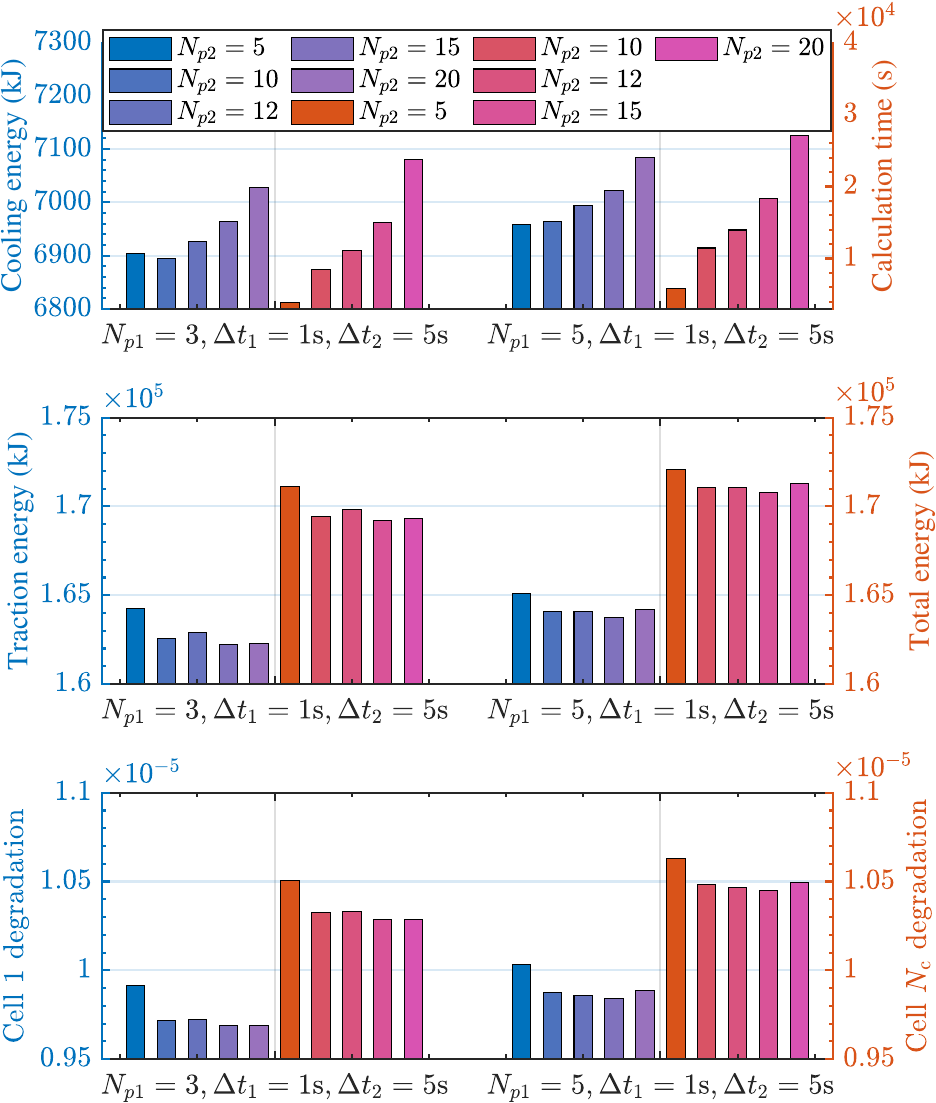}
    \caption{Sensitivity analysis on long horizon step of the proposed strategy.}
    \label{fig:MH_MPC_horizon step}
\end{figure}

Fig. \ref{fig:MH_MPC_horizon step} shows the results in terms of computation time, cooling energy, traction energy, and degradation of cell $1$ and cell $N_{\mathrm{c}}$. As the long prediction horizon $N_{\mathrm{p2}}$ increases, computation time rises significantly. Meanwhile, the traction energy and degradation of both cells are slightly reduced. Although increasing the long prediction horizon helps balance the trade-off between cooling energy and degradation, the marginal benefits in battery degradation reduction diminish as the prediction horizon lengthens. Moreover, the MH-MPC framework outperforms the SH-MPC framework, with improvements ranging from $6.35\%$ to $8.26\%$ in cooling energy consumption, $1.58\%$ to $2.76\%$ in traction energy, $11.12\%$ to $13.13\%$ in degradation loss for cell 1, $11\%$ to $12.84\%$ in degradation loss for cell $N_{\mathrm{c}}$, and $6.57\%$ to $8.92\%$ in battery degradation inconsistency, highlighting the superior performance of the MH-MPC framework in power and thermal management.

In conclusion, the MH-MPC configuration with $N_{\mathrm{p1}} = 3$, $N_{\mathrm{p2}} = 5$, $\Delta t_{1} = 1\mathrm{s}$, and $\Delta t_{2} = 5\mathrm{s}$ is suitable for practical application. This configuration under the MH-MPC framework achieves substantial improvements over the main benchmark: a $7.18\%$ reduction in computation time, $14.22\%$ in cooling energy consumption, $8.26\%$ in traction energy, $8.52\%$ in total energy, $22.47\%$ in degradation of cell 1, $23.42\%$ in degradation of cell $N_{\mathrm{c}}$, and 36.57\% in degradation inconsistency.

\section{Conclusion}\label{Conclusion}

In this paper, we propose an advanced IPTM strategy within the MH-MPC framework to enhance energy efficiency, ensure traffic safety and efficiency, regulate battery temperature, and reduce battery degradation for CAEVs. By analyzing the control and optimization problem, we reformulate it by focusing solely on the aging term, which generates advanced power and thermal management strategies that balance trade-offs among energy consumption, degradation, and temperature regulation over both short and long prediction horizons.
Specifically, the proposed strategy delivers notable improvements, reducing computation time by $7.18\%$, cooling energy consumption by $14.22\%$, traction energy consumption by $8.26\%$, battery degradation loss by over $22\%$, and degradation inconsistency by $36.57\%$ compared to the benchmark strategy. Additionally, we conduct sensitivity analyses to examine the influence of weighting parameters, sampling time, and prediction horizons on performance. The results demonstrate the potential for real-time implementation in extending battery lifespan and driving range while maintaining traffic safety and efficiency.

\ifCLASSOPTIONcaptionsoff
  \newpage
\fi

\bibliographystyle{IEEEtran}
\bibliography{refs}

\end{document}